\documentclass{article}

\usepackage{microtype}
\usepackage{graphicx}
\usepackage{subfigure}
\usepackage{booktabs} 

\usepackage{hyperref}

\usepackage[accepted]{mlsys2024}

\mlsystitlerunning{Compiler generated feedback for Large Language Models}

\begin{document}

\twocolumn[
\mlsystitle{Compiler generated feedback for Large Language Models}



\mlsyssetsymbol{equal}{*}

\begin{mlsysauthorlist}
\mlsysauthor{Dejan Grubisic}{meta,rice}
\mlsysauthor{Chris Cummins}{meta}
\mlsysauthor{Volker Seeker}{meta}
\mlsysauthor{Hugh Leather}{meta}
\end{mlsysauthorlist}

\mlsysaffiliation{meta}{Meta AI, Menlo Park, CA, USA}
\mlsysaffiliation{rice}{Rice University, Houston, TX, USA}

\mlsyscorrespondingauthor{Dejan Grubisic}{dx4@rice.edu}

\mlsyskeywords{Machine Learning, MLSys}

\vskip 0.3in

\begin{abstract}
We introduce a novel paradigm in compiler optimization powered by Large Language Models with compiler feedback to optimize the code size of LLVM assembly. The model takes unoptimized LLVM IR as input and produces optimized IR, the best optimization passes, and instruction counts of both unoptimized and optimized IRs. Then we compile the input with generated optimization passes and evaluate if the predicted instruction count is correct, generated IR is compilable, and corresponds to compiled code. We provide this feedback back to LLM and give it another chance to optimize code. This approach adds an extra 0.53\% improvement over -Oz to the original model. Even though, adding more information with feedback seems intuitive, simple sampling techniques achieve much higher performance given 10 or more samples.

\end{abstract}
]



\printAffiliationsAndNotice{}  

\section{Introduction}\label{submission}



\begin{figure*}[ht]
\begin{center}
\centerline{\includegraphics[width=\textwidth]{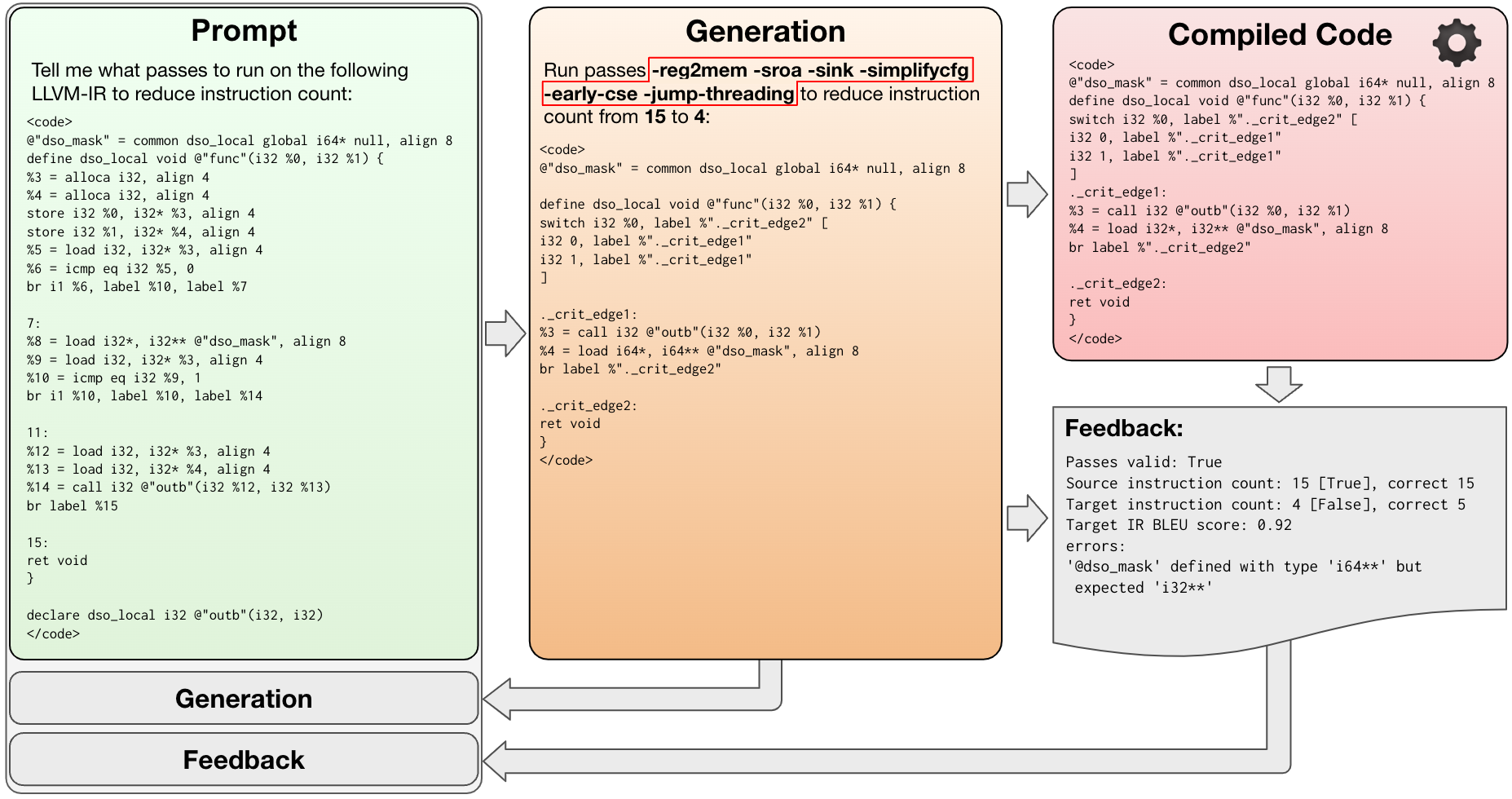}}
\caption{Feedback-directed model. First, we ask in the prompt LLM to optimize the instruction count of the given IR, then LLM generates the best optimization passes, instruction counts for starting and generated IR, and generated IR itself. Next, we compile the generated pass list and create feedback by checking if the generated pass list is valid, evaluating instruction counts, examining if the generated IR contains compilation errors, and calculating the BLEU score between the generated IR and the compiled IR. If some of the parameters of the feedback is problematic, we extend the original prompt with generation, compiled code, and feedback and ask it to try again. }
\label{fig/feedback_model}
\end{center}
\end{figure*}

\begin{figure*}[ht]
\begin{center}
\centerline{\includegraphics[width=\textwidth]{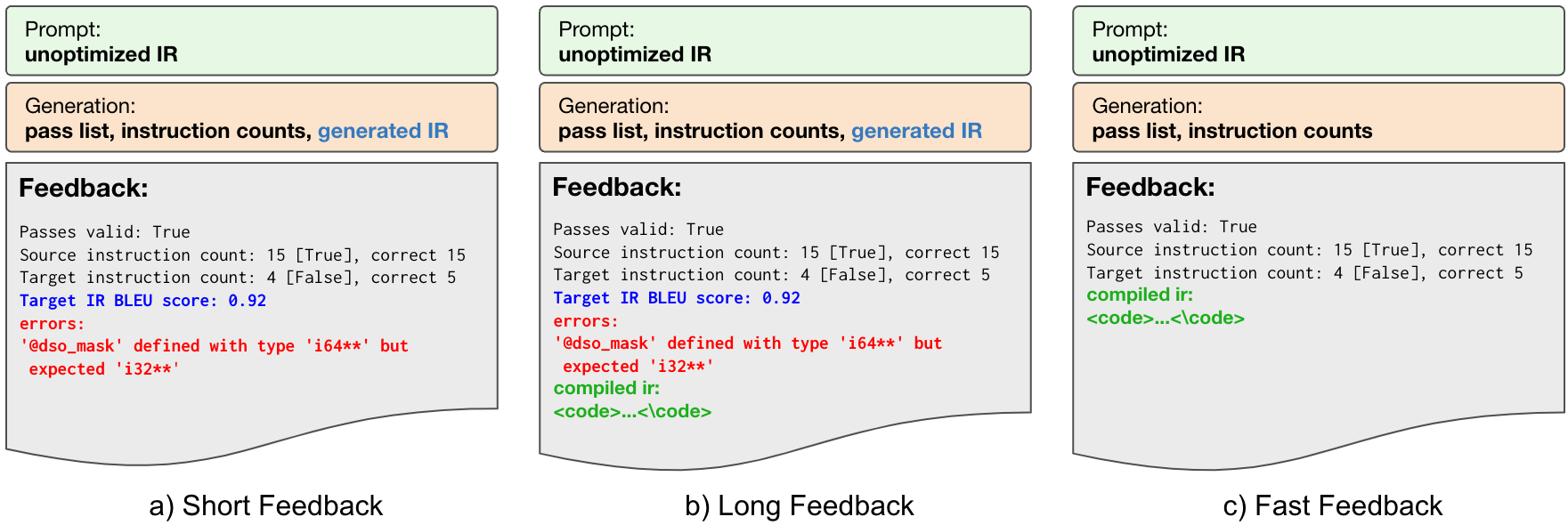}}
\caption{ Prompt structure of Feedback models. Short Feedback is the smallest in size and extends the prompt with just calculated metrics and error messages. Long Feedback contains the most information including compiled IR. Fast Feedback is the fastest to generate since it doesn't need the generation of IR to be calculated.}
\label{fig/feedbacks}
\end{center}
\end{figure*}

Large Language Models (LLMs) have proven their ability in the software engineering domain to generate the code and documentation~\cite{li_starcoder_2023, allal2023santacoder}, translate code between programming languages~\cite{transcoder, armengol2021learning}, write unit-tests~\cite{titanfuzz, schäfer2023adaptive}, as well as detect and fix bugs~\cite{ahmad2023fixing, xia2023automated}. The availability of large open-source code datasets~\cite{da2021anghabench, armengol2022exebench} and Github enabled models such as CodeLlama~\cite{llama-code}, ChatGPT~\cite{openai2023gpt4}, and Codex~\cite{chen_evaluating_2021} to develop a statistical understanding of various languages improving significantly the coding experience. Some of the models such as AlphaCode~\cite{li_competition-level_2022} are pretrained on competitive programming tasks which enables the model to optimize code on the source level for several languages.

Motivated by those successes we decided to embrace the use of LLMs further in compilers and optimize code on the level of LLVM IR. In contrast to optimizing source code, where LLM should provide suggestions and a programmer should accept it or not, in compilers, LLM is responsible for making the final decision. As a consequence, LLM is not allowed to change semantics in any case. To guarantee that syntax is not changed, we use LLM to suggest optimization strategies for the given code and direct the compiler to perform optimization passes. This way we can get the best of both worlds by navigating complex optimization space with LLM while keeping the code provably correct. 


Previous research in machine learning-driven code optimization has adopted various techniques, ranging from manually crafted features~\cite{mlgo, wang2018machine, ml4sysreview} to sophisticated graph neural networks (GNNs)~\cite{coreset, programl}. However, a common limitation across these approaches is that the representation of the input program provided to the machine learning model remains incomplete, resulting in some loss of critical information. For instance, in the case of MLGO~\cite{mlgo}, it leverages numerical features to offer insights into function inlining but falls short in faithfully reconstructing aspects such as the call graph and control flow. Similarly, PrograML~\cite{programl} employs graph-based program representations for GNNs, yet it overlooks essential details like constant values and certain types of information, hindering the faithful reproduction of program instructions.

To overcome this problem Cummins et. all~\yrcite{cummins2023llms-for-compilers} proposed LLMs for tuning LLVM optimization passes to reduce code size. By using LLVM IR code directly as an input, this approach demonstrated remarkable code reasoning abilities, outperforming default LLVM \textit{-Oz} optimization by 2.87\% while long-run autotuner used to generate data for LLM achieved 5\% improvement. Instead of generating just optimization passes the model also predicts the instruction count of source and target code and optimized IR which significantly increases the performance.

We extend ~\cite{cummins2023llms-for-compilers} further by evaluating the consistency of model generation by the compiler and providing feedback to the model. For each generation of the original model, we evaluate if the generated pass list is valid and if predicted instruction counts are correct and calculate bleu score~\cite{bleu} between the generated code and code we get by compiling the generated pass list. Additionally, we provide generated and compiled code in the feedback. 

We make the following contributions in this paper:
\begin{itemize}
    \item We present 3 compiler-generated feedback models for LLMs (Section \ref{sec:Feedback-directed LLMs})
    \item We evaluate 3 sampling methods with feedback (Section \ref{sec:Feedback model sampling})
    \item We evaluate iterative feedback generation (Section \ref{sec:Feedback model iterative algorithm})
\end{itemize}

With the Feedback method, we were able to increase the performance from 2.87\% to 3.4\% over the \textit{-Oz} with a single shot while failing to improve the performance of the original model when using sampling with more than 10 samples. Additionally, the Feedback method doesn't improve the performance of the original model when it is allowed to iterate 5 steps, while the original model uses sampling with the same number of inferences.

\section{Feedback-directed LLMs} \label{sec:Feedback-directed LLMs}
The performance of LLMs improves significantly when they are allowed to generate a series of reasoning steps to get to the final solution \cite{wei2022chain-of-thoughts}. This kind of behavior is particularly true for complex problems such as arithmetic, symbolic reasoning, and code generation. Additionally, increasing stochasticity and generating multiple different solutions, leads to superior performance. In both cases, the model often generates misguided solutions that impede their ability to reason correctly. Being able to get feedback on its generation and fix its errors could enable LLM to take one step further toward coming to a favorable solution.

We explore the model shown in Figure \ref{fig/feedback_model} as a three-step process. In the first step, the model starts from the prompt that contains only unoptimized IR and generates an optimization pass list, instruction count, and optimized IR itself. In the second step, we derive available metrics from generation with the help of the compiler and construct feedback. The feedback purpose is to quantify the consistency of generation and to point out where the internal model of the LLM diverges from the actual compiled IR. In the third step, we provide feedback to the model and give it a second chance.

To construct a feedback we evaluate if the generated pass list is valid, then we compile source IR with the generated pass list producing compiled IR. Next, we count the number of instructions of compiled IR and evaluate if the predicted source IR and optimized IR are correct. Since optimized IR could be derived from both generated IR and compiled IR, we save both metrics in the feedback.
Additionally, we validate if the predicted IR is compilable, save the error message if any, and calculate the Bleu score between the generated IR and compiled IR.

We compare 3 kinds of feedback (Figure \ref{fig/feedbacks}). Short Feedback 
contains predictions and answers for metrics and error messages. Long Feedback contains all derivable metrics and extends Short Feedback by Compiled IR. Since Short and Long Feedback both contain the metrics that come from generated IR, they require a full generation to be constructed. Fast Feedback avoids this by providing only metrics calculated from the pass list and instruction counts. This enables the model to stop generation early, terminating in just a few seconds which is about 10x faster than for other feedback.

When it comes to hardware efficiency, the process of appending Feedback data is extremely efficient. In this moment when the model generates the last output token, the GPU memory contains already prompt and generation. Adding Feedback would just be written in already allocated GPU memory, and the model would be ready for evaluation a second time.

Structuring the Feedback task after prompt and generation has one additional benefit. It reinforces learning of optimization tasks without feedback as well. This happens because the probability of one token depends only on the previous tokens and since we are appending a Feedback task after the task of optimization, it will not have an influence on it. This way we can use the same model for both optimization without feedback and with feedback.

Combining the Feedback approach with sampling can be an effective way of tuning applications. By increasing the temperature in LLM generation the model creates multiple strategies for optimization. Since this process is stochastic there is a higher chance there will be some errors in the generation. Learning the model and how to correct itself could enable it to develop the idea further by fixing itself until it comes to a promising solution.

\subsection{Feedback Metrics vs. Model Performance}

\begin{figure}[t]
\begin{center}
\centerline{\includegraphics[width=\columnwidth]{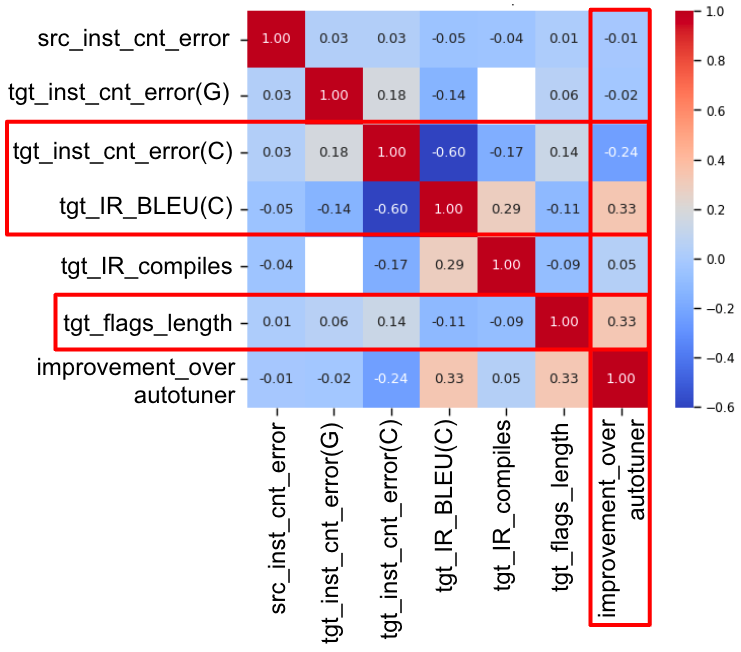}}
\caption{
    Correlation heatmap of metrics available at inference time. Input and output prompts are described with prefixes (src, tgt). Instruction counts are abbreviated with inst\_count. (G) stands for generation while (C) stands for compiled.
}
\label{fig/correlation_heatmap}
\end{center}
\end{figure}

We found that it is possible to derive a correlation between metrics available in the inference time and a model performance (Figure \ref{fig/correlation_heatmap}). There is a negative correlation between tgt\_inst\_cnt\_error(C) (the difference between predicted target instruction count and instruction count of IR got by compiling predicted passes) and improvement\_over\_autotuner. In other words, a smaller error in predicting the target instruction count means that the model is more likely to find a good solution. This is also the case if tgt\_IR\_BLEU(C) (generated IR compared to compiled IR bleu score) is high and the number of generated flags is large.

\begin{figure}[t]
\begin{center}
\centerline{\includegraphics[width=\columnwidth]{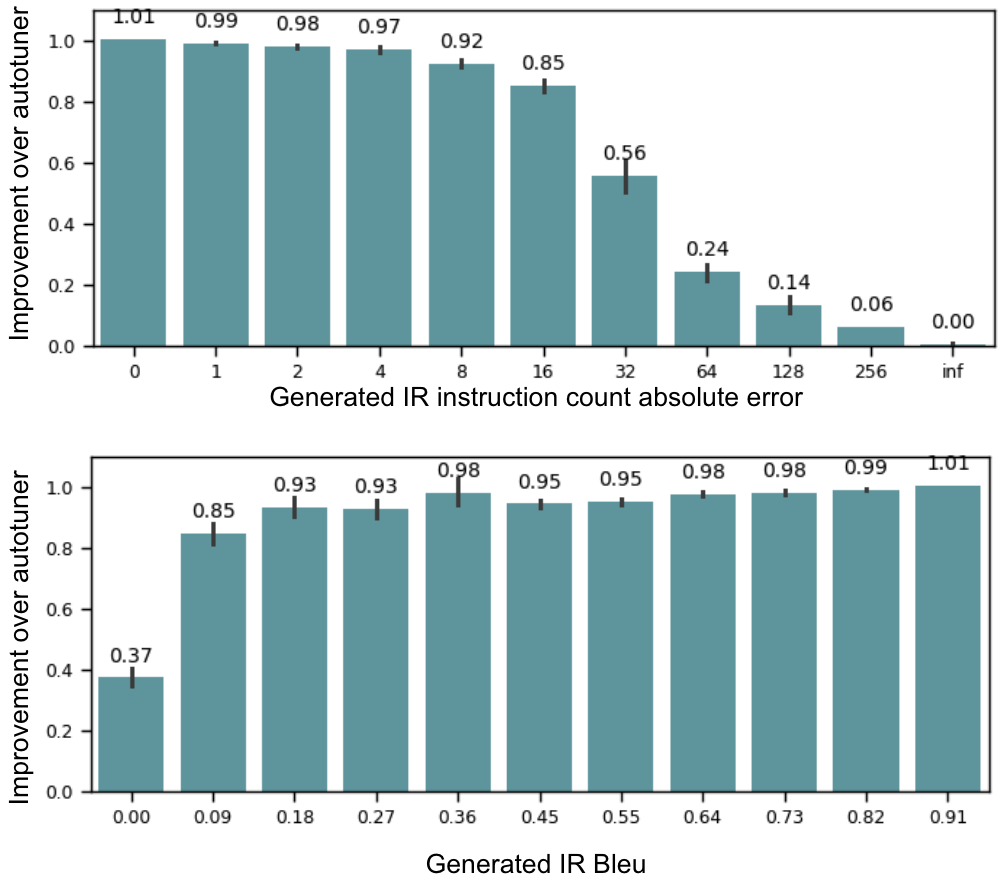}}
\caption{
    Distribution of absolute error in predicting optimized IR instruction count and Bleu score with respect to performance compared to autotuner.
}
\label{fig/correlation_bars}
\end{center}
\end{figure}

To understand more precisely the relation between tgt\_inst\_cnt\_error(C), tgt\_IR\_BLEU(C), and performance we plot the distribution of their values in Figure \ref{fig/correlation_bars}. When the model correctly predicts the target instruction count it also matches the performance of the autotuner. This means that when we detect this case in the inference, we can stop prompting and accept generated optimization passes. Similarly, we can stop prompting if the compiled and generated IR are equal, which results in tgt\_IR\_BLEU(C) being 1.

\section{The Model}
We train a 7B-parameter model with LLaMa 2 architecture \cite{llama} for each of the Feedback forms. As the starting point for training, we use the best checkpoint from \cite{cummins2023llms-for-compilers} which only predicts the best optimization passes for the given IR. We use the same Byte Pair Encoding \cite{bpe} tokenizer and model architecture that includes 32 attention heads, 4,096 hidden dimensions, and 32 layers, for a total of 7B parameters.

\begin{figure*}[ht]
\begin{center}
\centerline{\includegraphics[width=\textwidth]{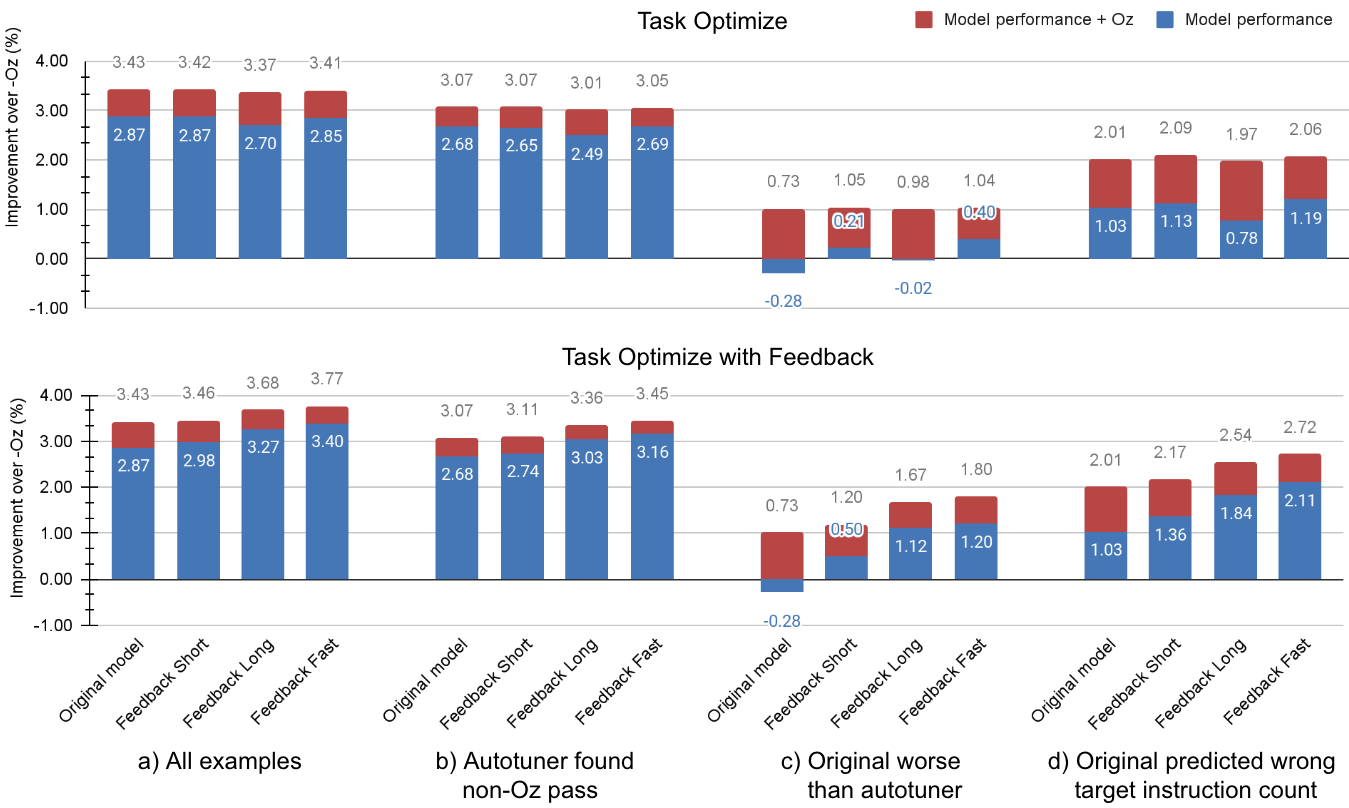}}
\caption{Comparison of the original and feedback models in reducing instruction count. The upper figure shows the performance of Task Optimize. The lower figure shows the performance on Task Feedback, where each model uses their format for feedback. Horizontally, we show the performance on all examples, examples where the autotuner's best pass is non-Oz, examples where the original model was worse than the autotuner, and examples where the original model mispredicted target instruction count. All the models keep the ability to perform Task Optimize while improving the performance when feedback is provided.}
\label{fig/models_comparison}
\end{center}
\end{figure*}


\subsection{Datasets}

We construct a training data set for each of the feedback forms. We get feedback data by evaluating the \textit{*best model} on 1 million training examples and 100 thousand test examples provided by \cite{cummins2023llms-for-compilers}. Additionally, we extract half of the examples from the test set to serve as a validation set.

For all forms of feedback, the prompt will have the structure described in Figure \ref{fig/feedbacks}. For expected generation, we keep the same format as the original work with an addition of the first line that indicates if the model is sure in its generation. Model outputs \textit{"I am sure!"} if the \textit{*best model} correctly predicted target instruction count, which is a strong indication that the model matched the performance of the autotuner. Otherwise, the model outputs \textit{"Let me try again."}.


\subsection{Training}
We trained all our models for 20,000 steps, with 64 A100 for about 60 GPU days. We use the AdamW optimizer~\cite{adamw} with $\beta_1$ and $\beta_2$ values of 0.9 and 0.95. We use a cosine learning rate schedule with 1,000 warm-up steps, a peak learning rate of $1e{-5}$, and a final learning rate of 1/10th of the peak. We used a batch size of 256 and each batch contains 786,432 tokens for Short and Fast Feedback and 1M tokens for Long Feedback for a total of 16B and 21B tokens respectively. The full training of 20,000 steps made 5.12 iterations over the training corpus.

\begin{figure*}[ht]
\begin{center}
\centerline{\includegraphics[width=\textwidth]{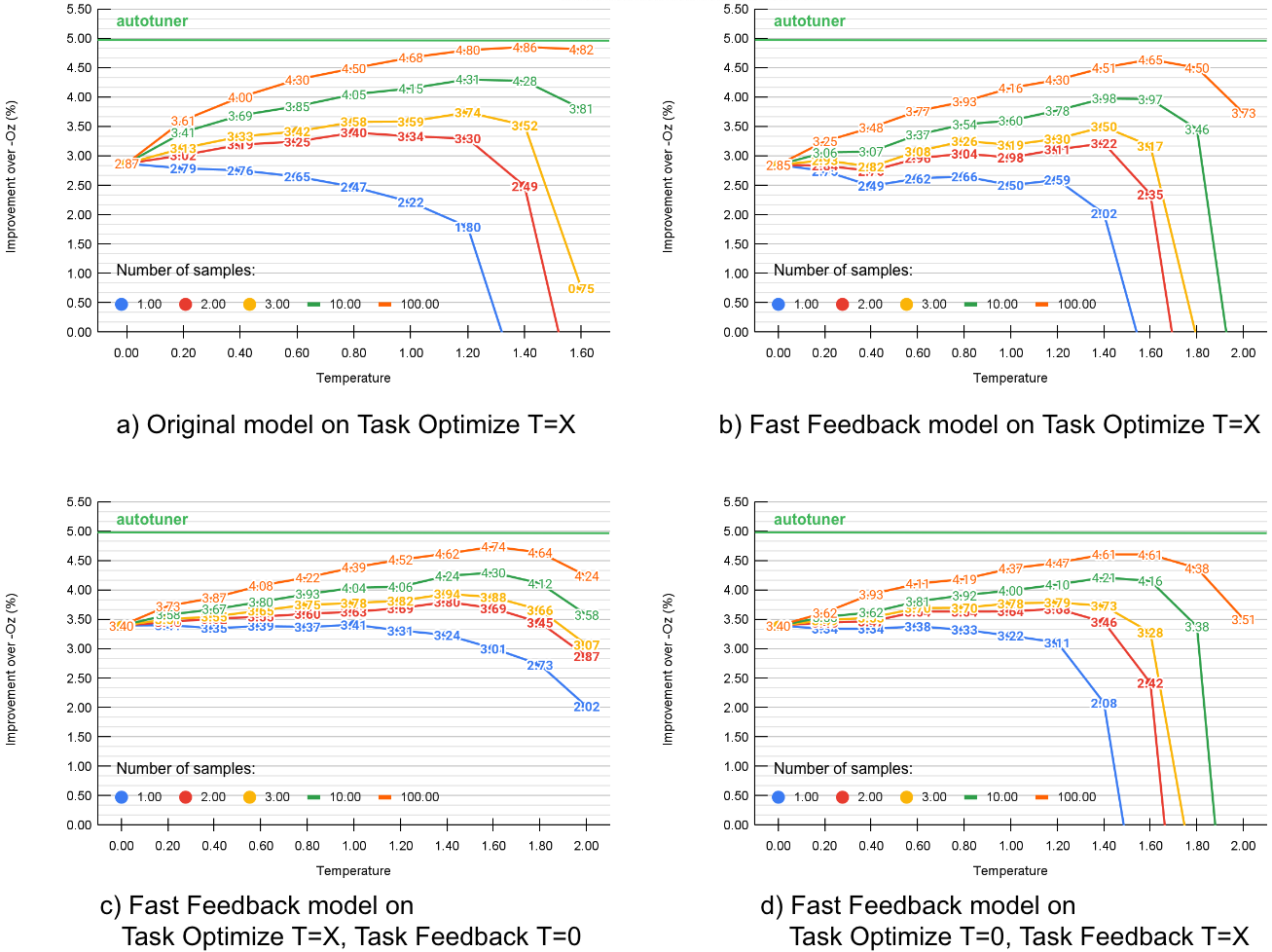}}
\caption{Sampling diagrams of the original and feedback models.}
\label{fig/sampling_plot}
\end{center}
\end{figure*}

\section{Evaluation}

In the evaluation, we are answering the following questions:
\begin{itemize}
    \item How does the feedback model compare to the original in Task Optimize and Task Feedback?
    \item How does the feedback model achieve when sampling is enabled?
    \item Can we use the feedback model to iteratively generate feedback and repair the current solution?
\end{itemize}

We found that the feedback model keeps the ability to optimize IR even without feedback. When it is allowed to apply two inferences, it can outperform the original model by 0.53\% closing the gap to the autotuner by 10\%. On the other hand, when the sampling is enabled, we show that the original model achieves up to 98\% of the autotuner performance given 100 samples. We evaluate 3 sampling strategies for the feedback model and show that they all fail to match the sampling of the original model. Finally, we compare the performance of the iterative feedback model with the original model given the same amount of computation per sample and we show that with 2 or more samples and a temperature higher than 0.4, the original model outperforms the feedback model.

\subsection{How does the feedback model compare to the original in Task Optimize and Task Feedback?}

We compare all three Feedback models with the original on the Task Optimize and Task Feedback (Figure \ref{fig/models_comparison}). In Task Optimize the input prompt consists only of the input IR, while in Task Feedback each model will append the input prompt with the feedback they got from the previous generation in the format defined in Figure \ref{fig/feedbacks}. Additionally, we show performance on all examples, examples where the autotuner found a non-Oz optimization pass, examples where the original model was worse than the autotuner, and examples where the original model mispredicted the instruction count. Furthermore, we show the performance of the model combined with \textit{-Oz}.

All of the Feedback models perform similarly on average to the original on Task Optimize even without being trained on that task explicitly in the feedback finetuning. Moreover, the feedback models even improved the performance for the examples where the original model performed worse than the autotuner by 0.6\% for Fast Feedback. The reason for this is that we add extra information to the input, which enables the model to discriminate hard examples and learn them easier.

In the plot below, we feed the output from Task Optimize to each Feedback model and apply Task Feedback, while keeping the results from Task Optimize for the original model. All the Feedback models improve the performance of the original model by 0.19\% with Short Feedback, 0.4\% with Long Feedback, and 0.53\% for Fast Feedback. Most of the improvement comes from the examples where the original model performed worse than the autotuner and the examples where the model mispredicted the generated instruction count. Here the Fast Feedback model outperforms the original model by 1.48\% and 1.07\% respectively.

It is interesting that the Fast Feedback model performs better than Long Feedback, in spite of using a subset of information in the input prompt. We believe the reason for this is that using generated IR in the input prompt introduces noise and makes the model find the relevant information. Since the Fast Feedback model doesn't need to generate IR to create feedback we can iterate with it much faster. For the next experiments, we use the Fast Feedback model for further evaluation.

\subsection{Feedback model sampling} \label{sec:Feedback model sampling}
We compare the Fast Feedback model against the sampling of the original model with different temperatures (Figure \ref{fig/sampling_plot}). In the first two plots, we show the original and Fast Feedback model on Task Optimize with different temperatures. The original model achieves the peak performance of an astonishing 98\% of the autotuner given 100 samples on the temperature 1.4. This result demonstrates the significant power of sampling that opens up interesting directions for future research. With 10 samples the original model achieves about 86\% of the autotuner which is a significant win compared to 57\% of the temperature 0.

The third plot evaluates the Fast Feedback model starting from Task Optimize with given temperatures and applying Task Feedback with temperature 0. The Fast Feedback model achieves up to 93\% on Task Optimize on the temperature 1.6. This is expected since it was not finalized on Task Optimize. Interestingly, it has the peak for the higher temperature, which indicates that its logit distribution is sharper initially and it is more conservative in choosing the next tokens.

The fourth plot evaluates the Fast Feedback model starting from Task Optimize on temperature 0 and applies Task Feedback with given temperatures. For the lower number of samples (1, 2, 3) it achieves higher performance than the original model, but it doesn't keep the advantage for 10 or more samples. The reason for this might be that the Fast Feedback model was trained on the feedback of the original model with temperature 0, and the generations become significantly different when the temperature is on. This indicates that training the Feedback model on generations with temperature can increase the performance.

Finally, sampling the Fast Feedback model after the Task Optimize with the temperature 0 fails to deliver performance as well. Similar to the previous method, this method performs well for a smaller number of samples, while it even has lower results than Task Optimize with 100 samples. This might indicate that it generates more conservative solutions than the original model.

\begin{figure}
\begin{center}
\centerline{\includegraphics[width=\columnwidth]{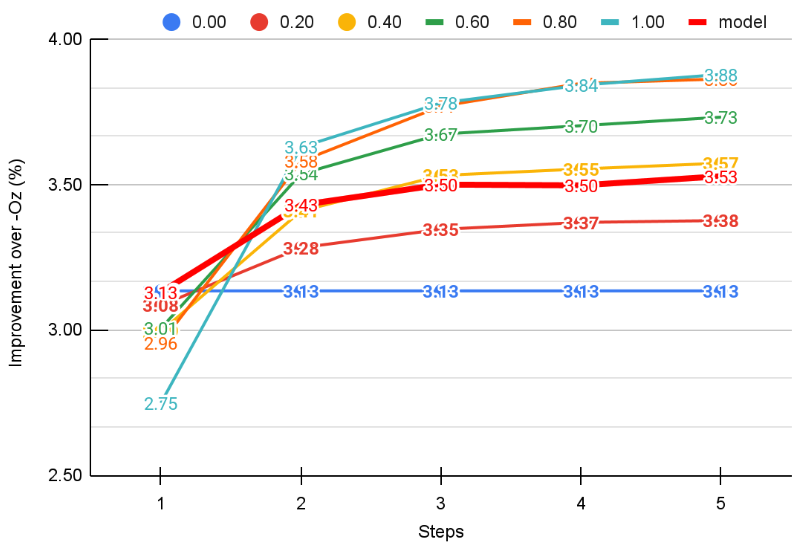}}
\caption{
    Comparison of the iterative approach (model) versus the sampling of the original model with the same amount of computation. In each step the Fast Feedback model generates feedback for the next step, applying Task Optimize in the first step and Task Feedback afterwards. Once the model outputs "I am sure!" we stop. We allow the same number of generations for the original model.
}
\label{fig/bridge_steps}
\end{center}
\end{figure}

\section{Additional Experiments}

\subsection{Feedback model iterative algorithm}\label{sec:Feedback model iterative algorithm}
In this section, we evaluate the capability of the fast feedback model to iterate based on previous solutions and compare it to the sampling of the original model with the same number of inferences. First, the Fast Feedback model applies Task Optimize, which generates feedback. Then the model applies Task Feedback iteratively by using the feedback of the previous task and generates the feedback for the next step for a total of 5 steps. After every generation, we check if the Fast Feedback model generated \textit{"I am sure!"} and use that as the final result. For the original model we applied Task Optimize and sampled each example as many times as we made steps with the feedback model We use temperatures in the range [0,1] and show the cumulative performance of each approach in Figure \ref{fig/bridge_steps}. 

The Fast Feedback model has a similar performance to the original model with temperature 0 on Task Optimize while making the most of performance improvements in the second step (Task Feedback), and slowly increasing by iterating on Task Feedback, achieving performance similar to the original model with temperature 0.4. The original model with temperature 1 starts with the lowest performance, but outperforms all other approaches from step 2. This result demonstrates that for our problem the sampling is more powerful than iteration and should be used instead.

\section{Related Work}

In recent years we have witnessed tremendous progress in the development of Large Language Models for software engineering. These models are able to generate code and documentation~\cite{openai2023gpt4, llama-code, chowdhery2022palm, li_starcoder_2023, allal2023santacoder, fried_incoder_2023, gunasekar2023textbooks}, translate code between programming languages~\cite{transcoder, armengol2021learning, transcoder-ir}, write unit-tests~\cite{ye_2021, titanfuzz, schäfer2023adaptive}, as well as detect and fix bugs~\cite{ahmad2023fixing, xia2023automated}.

The availability of various open-source code datasets~\cite{da2021anghabench, armengol2022exebench} and the accessibility of platforms like GitHub have enabled models such as CodeLlama~\cite{llama-code}, ChatGPT~\cite{chatgpt}, and Codex~\cite{chen_evaluating_2021} to elevate their coding capabilities. However, it's important to note that these models were not explicitly designed for code optimization. For instance, ChatGPT can perform minor optimizations, like tagging variables for storage in registers, and even attempt more substantial improvements such as vectorization. Nevertheless, it often encounters confusion and makes errors, leading to incorrect code outcomes.

On the other hand, models such as AlphaCode~\cite{li_competition-level_2022} generate a performant solution by optimizing code on the source level. AlphaCode is fine-tuned on competitive programming problems from the Codeforces platform while using 715.1 GB of code data from GitHub for pretraining. Furthermore, it generates a large corpus of potential solutions from which it chooses the 10 best solutions by implementing sophisticated clustering and filtering mechanisms. Similar to Alphacode we demonstrate the power of sampling while targeting compiler-level optimizations instead of the optimizations on the source code.

When it comes to fundamental LLM improvements, Wei et. all~\yrcite{wei2022chain} showed that significant improvement of the LLM can be obtained by splitting answers in step by step manner for models with more than 10B parameters. Brown et. all showed that a few-shot prompting~\yrcite{brown2020language} based on extending prompt with similar (question, answer) pairs additionally increases the performance of the model. Yang et. all~\cite{yang2023large} extend this approach further by iteratively adding generated solutions and their evaluations to the original prompt, together with few-shot prompting. In our approach, we provide more informative feedback based on inference time evaluation of model generations that includes validation of generated pass list, evaluation of predicted instruction counts, and optimized IR.

Finally, the problem of compiler pass ordering has been explored for many decades, as indicated by prior works~\cite{bodin1998iterative, kisuki_combined_2000, fursin_evaluating_2005}. In recent years, machine learning has become an avenue for this optimization task, with various approaches proposed~\cite{wang2018machine, ml4sysreview, coreset, autophase, agakov_iterative_2006, ogilvie_minimizing_2017}. It's noteworthy that the implementation of machine learning in compiler technology extends beyond just optimizing pass order and encompasses a wide range of other challenges~\cite{mlgo, Ashouri_2022, hajali2020neurovectorizer, cummins_end2end,phothilimthana2021flexible}. We extend these efforts further by pioneering the use of Large Language Models in compiler optimization.

\section{Limitations and Future Work}
Sampling has great potential to improve the performance of the original model. In this paper, we saw that the Fast Feedback model fails to improve the performance of the original model when there are larger than 10 samples. This indicates either that the Fast Feedback model doesn't know how to deal with generations on higher temperature, or that it generates always the same output. To mitigate this, we can generate a dataset by sampling the original model and train the Fast feedback model on these data.

Another approach is to disregard the feedback component and implement a smarter sampling heuristic for the original model. For this, we can use beam search and similar techniques when sampling the tokens. AlphaCode~\cite{li_competition-level_2022} for example generates many possible solutions and then uses clustering and filtering techniques to select the best answers. Developing methods in this area could be more generally applicable in the LLM community.

\section{Conclusions}

In this paper, we introduce the idea of compiler-generated feedback for Large Language Models. In our approach, the model starts from unoptimized LLVM IR and predicts the best optimization passes and instruction counts and optimized IR. Then we construct feedback on the model's generation and ask the model to try again. We explore 3 formats of feedback with various degrees of information. All forms of the feedback outperform the original model on temperature 0 by 0.11\%, 0.4\%, and 0.53\%.

We further explore the sampling properties of the Fast Feedback and compare it to the original model. The Fast Feedback model fails to improve the performance of the original model when 10 or more samples are given. Additionally, we compare the sampling of the original model to the iterative generation of the Feedback model with the same amount of computing. In this case, we found that sampling of the original model between temperatures [0.4, 1] achieves higher performance. 

From this, we conclude that sampling is an irreplaceable technique for getting high performance and more research needs to be done. Potential future directions could be training the Fast Feedback model by collecting the feedback of the original model sampled on the given Temperature. This way Feedback model should become more capable of optimizing more fuzzy generations. Another approach would be to explore further sampling of the original model and develop more sophisticated algorithms that preserve diversity without generating incoherent text.

\bibliography{example_paper}
\bibliographystyle{mlsys2024}



\end{document}